# Grating lobe reduction in plane wave imaging with angular compounding using subtraction of coherent signals

Zhengchang Kou, *Member, IEEE*, Rita, J. Miller, and Michael L. Oelze, Senior *Member, IEEE*

*Abstract*—Plane wave imaging (PWI) with angular compounding has gained in popularity over recent years because it provides high frame rates and good image properties. However, most linear arrays used in clinical practice have a pitch that is equal to than the wavelength of ultrasound. Hence, the presence of grating lobes is a concern for PWI using multiple transmit angles. The presence of grating lobes produces clutter in images and reduces the ability to observe tissue contrast. Techniques to reduce or eliminate the presence of grating lobes for PWI using multiple angles will result in improved image quality. Null subtraction imaging (NSI) is a nonlinear beamforming technique that has been explored for improving the lateral resolution of ultrasonic imaging. However, the apodization scheme used in NSI also eliminates or greatly reduces the presence of grating lobes. Imaging tasks using NSI were evaluated in simulations and physical experiments involving tissue-mimicking phantoms and rat tumors *in vivo*. Images created with NSI were compared with images created using traditional delay and sum (DAS) with Hann apodization and images created using a generalized coherence factor (GCF). NSI was observed to greatly reduce the presence of grating lobes in ultrasonic images, compared to DAS with Hann and GCF, while maintaining spatial resolution and contrast in the images. Therefore, NSI can provide a novel means of creating images using PWI with multiple steering angles on clinically available linear arrays while reducing the adverse effects associated with grating lobes.

## I. INTRODUCTION

Improving the quality of ultrasonic B-mode imaging remains an important area of research because B-mode imaging is ubiquitous in the clinical setting. In general, image quality is a function of contrast, spatial resolution and the signal-to-noise ratio (SNR). Different approaches have been explored to improve each of the image quality features often by trading off one quality to improve another. For example, approaches have been applied to improve the SNR of B-mode imaging using techniques like coded excitation and pulse compression [1], [2] or compounding [3]. Techniques to improve spatial resolution in B-mode ultrasound have been attempted using a variety of methods including beamforming techniques [4], coded excitation [5] and signal processing [6]. Finally, Contrast to noise ratio (CNR) is often considered the most important image quality metric. Because the contrast of ultrasound targets is often inherently low, it is important to remove artifacts that can reduce the ability to observe contrast in B-mode imaging. Reverberation and clutter due to things like side lobes or grating lobes can result in a reduction in the observable contrast of targets in B-mode ultrasound. Therefore, a large number of studies have occurred and papers have been written about reducing the impact of reverberation and clutter on B-mode imaging.

The list of works pertaining to the goal of improving contrast in B-mode ultrasound is very long and calling attention to all of the works in the field is the role of a review paper. However, in this work we focus on techniques to reduce side lobe and grating lobe artifacts in B-mode imaging when applying plane wave image (PWI) compounding. The idea of PWI coherent compounding was first introduced in a series of papers by Lu [7]-[12] and then popularized by Montaldo et. al. [13]. In PWI in 2D, a linear array sends out "plane" wave transmissions at multiple angles by firing off each element of the array simultaneously for zero-degree (or broadside) plane waves or with sequentially increasing or decreasing time delays to elements resulting in planes wave transmissions at different angles relative to broadside. Image quality is improved by coherently or incoherently compounding received signals from plane wave transmissions from multiple angles resulting in the ability to focus throughout the field.

One of the main attractive features of PWI is that it facilitates ultrafast B-mode imaging with frame rates on the order of $10^3$ per second or higher [14]. PWI and ultrafast imaging have been instrumental in improving vasculature imaging [15]-[17]. Further, PWI and ultrafast imaging have allowed the development of super resolution imaging for visualizing the vasculature for cardiac applications [18], [19], tumor imaging [20], [21] and functional imaging of the brain [22], [23]. Therefore, the adoption of PWI has had a significant impact on ultrasonic imaging with many of these techniques making their way into the clinic.

This work was supported by the National Institutes of Health under Grant R21EB024133 and R01CA251939. (*Corresponding author: Michael L. Oelze*).

The authors are with the Department of Electrical and Computer Engineering, Beckman Institute for Advanced Science and Technology, University of Illinois at Urbana-Champaign, Urbana, IL 61820 USA (email: zkou2@illinois.edu; oelze@illinois.edu)

One issue associated with using PWI on most clinically available linear arrays is that these linear arrays were not constructed with the steering of beams at moderate to high angles in mind. Most linear arrays have a pitch that is close to the wavelength of the center frequency of the probe and a few arrays exist where the pitch is larger than a wavelength. For broadside, i.e., zero-degree angle plane wave, beam formation, such a pitch would result in appreciable grating lobes in the field of view [24]. As the plane wave is steered away from broadside, grating lobes become more obvious in the field of view. For broadband probes, these grating lobes are reduced in magnitude compared to the main lobe but their energy is smeared out spatially. The high amplitudes of these grating lobes result in image artifacts and can greatly reduce the observable contrast of targets in the imaging field and potentially obscure important image details.

To address grating lobes in PWI, several approaches have been explored. In one approach, nonuniform plane wave angles were used to help suppress grating lobes [25]. In simulation, this technique reduced the amplitude of some grating lobes but did not completely mitigate their presence. In another approach, physically shifting of the transducer laterally between successive frame acquisitions was used to decrease the effective pitch between elements and mitigate grating lobes [26].

Like grating lobes, side lobes also produce clutter in ultrasound B-mode images. Reducing side lobe levels of ultrasonic beams has been an area of research since the foundations of diagnostic ultrasound. The simplest method for reducing side lobe levels in array imaging is to implement apodization using tapered windows. By applying a tapered apodization, such as a Hann apodization, the side lobe levels can be reduced but at the expense of a broadening main lobe [27]. Many other approaches to reducing side lobe levels have been investigated over the years with different levels of success [27-41]. However, for many of these techniques success in side lobe reduction comes with a tradeoff of higher computational cost.

Recently, our group introduced a nonlinear beamforming technique, null subtraction imaging (NSI), that utilized three apodization schemes on receive only [4]. In this technique, the first apodization consisted of a subaperture of $N$ elements where the first $N/2$ elements had a +1 weight and the second $N/2$ elements had a -1 weight resulting in a zero mean apodization across the receive subaperture. The resulting receive beam pattern had a null at zero degrees. The second apodization was equal to the first apodization but with a small DC offset value added to the apodization resulting in a nonzero mean apodization across the receive subaperture. The third apodization was a flipped version of the second aperture. The envelopes of the received and beamformed signals for the second and third apodizations are then summed and the envelope for the first apodization is subtracted from the sum. This results in the zero-degree null becoming the beam and accompanied by low side lobe levels.

In this study, we examine the use of NSI to dramatically reduce grating lobes and side lobes when using PWI and both coherent and incoherent compounding techniques. To test the approach, we constructed images of wire targets and acquired images from tissue-mimicking phantoms and from rat tumors *in vivo*. We compared the NSI approach to Hann apodization and to a generalized coherence factor (GCF) approach [31]. Grating lobe levels were assessed relative to the main lobe energy and image quality was assessed through reduction in clutter energy in the *in vivo* images.

## II. METHODS

*A. NSI*

NSI makes the use of multiple apodizations on receive on a linear array to create multiple images and then subtracts combinations of the images from each other to obtain a composite image with improved properties in terms of sidelobe levels, grating lobe levels and main lobe width. Specifically, transmission of ultrasound can occur through multiple means, e.g., linear sequential scanning with focusing or plane wave coherent compounding. However, linear sequential receive beamforming on receive is required for NSI implementation. In linear sequential scanning, subapertures are used to beamform and create individual scan lines, then the subapertures are electronically translated to create the next scan lines. An image is formed by combining these scan lines and possibly interpolating lines in between the actual measured scan lines.

With NSI, three images are created from the scan lines by applying three different apodizations on each subaperture [4]. The first apodization consists of a zero-mean window with the first half given the value of +1 and the second half given the apodization value of -1,

$$A_{ZM,i} = \begin{cases} 1, & 1 \leq i < \dfrac{N}{2} \\ -1, & \dfrac{N}{2} \leq i < N \end{cases},$$

where $A_{ZM,i}$ is the zero-mean apodization and $N$ is the number of elements in the subaperture. The second apodization is the same as the zero-mean apodization except for a DC offset,

$$A_{DC1,i} = A_{ZM,i} + c$$

and the third apodization is a flipped version of the second. An NSI image is created by subtracting the zero-mean image from the sum of the dc images,

$$E_{NSI} = \frac{E_{DC1} + E_{DC2}}{2} - E_{ZM}$$

where $E_{ZM}$ is the zero-mean envelope image and $E_{DC1}$ and $E_{DC2}$ are the two DC offset envelope images. The resulting B-mode image is constructed by converting to a decibel scale and scaling the image to the maximum value (i.e., 0 dB) and displayed in grayscale,

$$E_{\text{NSI,dB}} = 20\log_{10}[E_{\text{NSI}}] - \max\{20\log_{10}[E_{\text{NSI}}]\}.$$

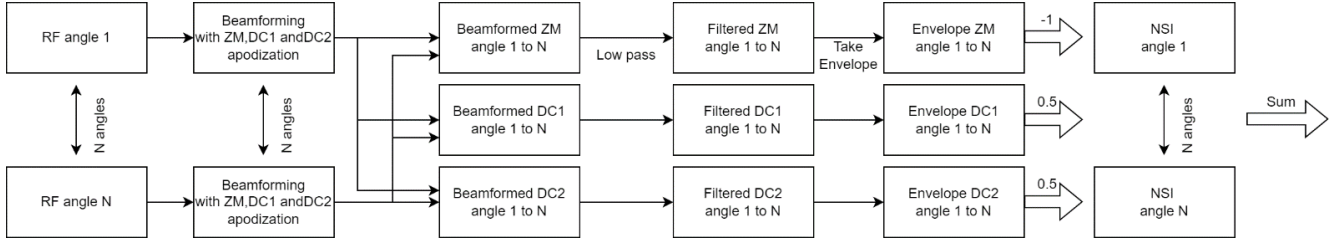

**Fig.1. Block diagram of IC-NSI. The raw channel RF data is first beamformed with three different apodizations for each angle. A LPF is applied to each apodization result in the angular dimension. The envelopes of the filtered ZM signal are subtracted from the sum of the filtered DC signal. Finally, incoherent compounding is performed on all the angular data to generate the final image.**

It has been shown that NSI could effectively improve the resolution by reducing main lobe width [4].

However, NSI also reduces the grating lobe levels. The reason for this reduction is based on the broadband nature of ultrasonic imaging. For the zero mean apodization, a null exists at 0°. On the other hand, grating lobes will not have a null because, due to the broadband nature of ultrasonic imaging, only a few sequential elements in a subaperture will contribute to the construction of the grating lobe at a single point in the field. Hence, the grating lobes will not see a zero-mean aperture. Similar grating lobe profiles will be produced in the 2nd and 3rd apodizations. Therefore, similar to the side lobes, the grating lobes are cancelled when subtracting the zero-mean apodization image from the DC offset images.

*B. Filtered Incoherent NSI*

In this work, we also propose a filtered incoherent NSI (IC-NSI) as an imaging scheme that is designed to maintain the advantages of both the higher lateral resolution of NSI and the better CNR of conventional methods for forming B-mode images. The IC-NSI process is illustrated in Fig.1. The raw RF data are first beamformed with three NSI apodizations individually for each angle. Then, the beamformed RF data are filtered. An Equiripple FIR LPF with normalized passband frequency at 0.3, stopband frequency at 0.8 and stopband attenuation at 60 dB was generated by Matlab Filter Designer and applied to the angular dimension of the beamformed data individually for each NSI apodization. After filtering, the envelopes were detected for each angle and each apodization. Then, the ZM image was subtracted from the sum of DC images individually for each angle. The resulting NSI images for each angle are summed together to generate the final image. In this way, the lateral resolution of NSI is maintained by the LPF, which filters out the signals that vary rapidly over different steering angles. At the same time a smoother speckle pattern is preserved for incoherent compounding in comparison to coherently summed NSI (C-NSI) [4]. Application of the LPF to the C-NSI did not result in any appreciable differences.

III. EXPERIMENTAL SETUP

*A. Simulation*

To evaluate the grating lobe reduction on imaging performance, we first simulated imaging of single targets using the Field II [43],[44] simulator. In the simulation we placed a single scatterer in the field of view. The simulation parameters were set according a L14-5/38 array transducer (Ultrasonix, BC, Canada). This particular array was used in the animal studies because of its broad bandwidth (nominal 5 to 14 MHz) and was, therefore, used in simulations to match the physical experiments. The array had a nominal center frequency of 7.82 MHz ($\lambda \approx 197$ µm), 128 elements, a pitch of 0.3048 mm, total length of 38 mm and an elevational focus of 16 mm. A fixed F-number of 1.5 was used in beamforming process. The Field II simulator represents an ideal case where all elements on the array are assumed to be perfectly matched in their element factors and sensitivities and the SNR can be controlled. A wire target at a depth of 5 mm was imaged using PWI with a total of 33 plane waves transmitted with steering angles spanning from -16 to +16 degrees in 1 degree increments. With this pitch, grating lobes are predicted to occur at angles close to +/- 40°.

*B. Physical Samples*

In physical experiments, grating lobe reduction was assessed using NSI for different imaging tasks from three different types of samples: wire targets in water, a tissue-mimicking phantom and rat tumors *in vivo*.

*Wire Target Experiments*

In the wire target experiments, a 100 $\mu$m diameter tungsten wire was placed in a tank of degassed water, which was then scanned by the linear array. A total of 33 plane waves were transmitted with steering angles spanning from -16 to +16 degrees in 1-degree increments. Received signals were recorded by Verasonics system on all 128 channels. On receive, a fixed f-number subaperture was used to create each scan line using delay and sum (DAS). The usable subaperture size is limited by the total number of elements (channels) when the depth is too large. When the subaperture moved beyond the edges of the array, we padded zeros to span the expected aperture. Images were created using three approaches: simple DAS beamforming with a Hann apodization, the same but with the GCF applied and using NSI. For NSI, different values of the DC offset were chosen (0.1 and 1.0) to quantify grating lobe level reduction versus the spatial resolution gain provided by NSI, i.e., a smaller DC offset results in better lateral resolution and lower grating lobe levels. In addition, NSI was assessed when using either coherent summing of the plane wave (C-NSI) or incoherent summation (IC-NSI). For GCF, the cutoff frequency $M_0$ [31] was empirically chosen to be 2 because it provided the best contrast-to-noise ratio (CNR) in the phantom contrast targets.

*CIRS Phantom Experiments*

In the phantom experiments, we scanned the CIRS 040GSE phantom (Computerized Imaging Reference Systems, Norfolk, VA, USA). As before, 33 plane waves were fired with steering angles equally spanning from -16 to +16 degrees. Received signals were recorded by Verasonics system on all 128 channels. Images were constructed using the same beamforming approaches that were performed in the wire target experiments.

*In Vivo Experiments*

All animal experiments were approved by the Institutional Animal Care and Use Committee at the University of Illinois at Urbana-Champaign. Tumors were induced in the mammary fat pad of a female F344 rats by injecting MAT tumor cells (5×10$^2$cells in 100μL) on each side of the abdomen. Once the tumors grew to 5−15mm in diameter, the animals were anesthetized using isoflurane and imaged. The skin above a tumor was shaved and tumors were coupled to the transducer array using ultrasound gel. Seven plane waves were transmitted with steering angles equally spanning from -12 to +12 degrees. All signals were received using the Verasonics system and the same beamforming approaches used in the wire target experiments were also used in the tumor imaging.

C. Image Quality Metrics

To evaluate the grating lobe reduction performance, we compared the lateral profile of axially integrated power estimated from the wire targets using Hann apodization and NSI because the energy in the grating lobes spread axially. We also estimated the amplitude differences between the grating lobes and main lobe for Hann apodization and NSI. Contrast to noise ratio (CNR) was measured for the CIRS phantom scan data. However, for the tumor images, the improvement in image quality was estimated by quantifying the reduction in the image intensity artifact due to grating lobes. Specifically, the intensity in regions observed to be artifacts from the grating lobes were compared between images created using the different beamforming approaches.

III. RESULTS

A. Simulation

Images from a single scatterer using Hann apodization, the three NSI's apodization (DC offset of 1.0) and the combined NSI image based on single zero-degree plane wave simulations are shown in Fig. 2. The grating lobes appear as smeared out intensity blobs between 5 and 7 mm axially and about ±4 mm laterally from the scatterer (approximately 40°). The Hann image shows that the grating lobes are symmetric about the scatterer. The DC offset images show grating lobes only on the left or right depending on which apodization is displayed. The zero-mean apodization is similar to the Hann apodization except for the null at broadside, i.e., zero degrees. Note that the grating lobes produced by the zero-mean apodization do not have a null at the center. In the NSI image, the grating lobes are subtracted out.

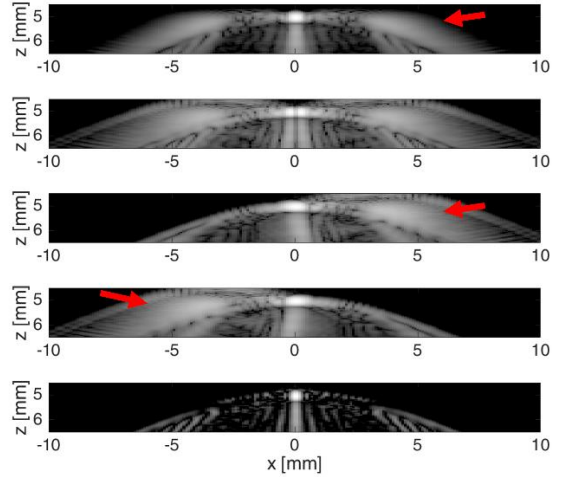

**Fig.2. Top to bottom: Single scatterer image using Hann apodization, using the zero mean apodization, the zero mean plus DC offset of 1.0, the flipped version of the previous apodization, and the NSI image. Dynamic range is 60 dB for all five images.**

The lateral profile of axially integrated power for the different approaches basing on a single zero-degree plane wave simulation are shown in Fig. 3. The beamwidths from the NSI are narrower as the DC offset is decreased. Side lobe levels are also greatly reduced when using NSI compared to Hann apodization. Finally, grating lobe levels for the Hann apodization are at about -20 dB down from the main lobe. However, for NSI the grating lobes are below -60 dB from the main lobe level.

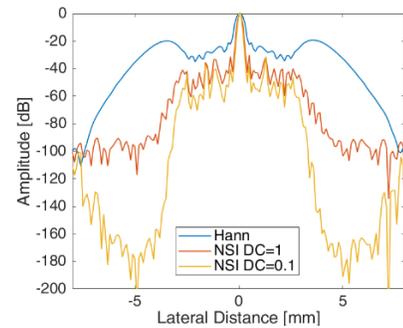

**Fig.3. Lateral profile of axially integrated power from Hann and NSI with different dc offset resulting basing on Field II single scatterer simulation.**

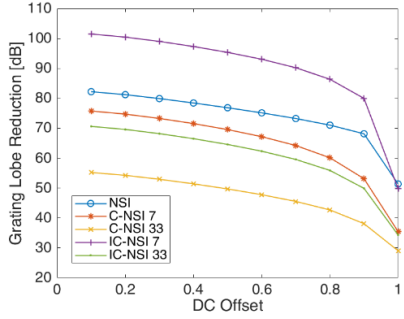

Fig.4. Grating lobe reduction of C-NSI and IC-NSI compared to Hann apodization estimated from axially integrated power of simulations with a single scatterer when using different numbers of angles for compounding.

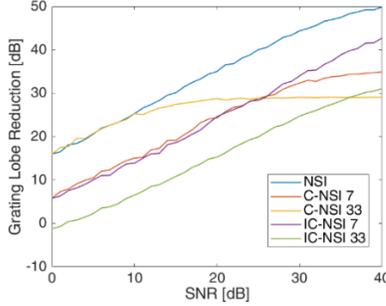

Fig.5. Grating lobe reduction of NSI compared to Hann apodization estimated from axially integrated power of simulations with a single scatterer under different input raw RF channel data SNR.

Using the Field II simulator and a single scatterer, the reduction in the grating lobes was quantified for different DC offset values. The reductions in grating lobe levels for C-NSI (coherently summed) and IC-NSI (incoherently summed) with different DC offset and different number of angles used for compounding are shown in Fig. 4 with respect to Hann apodization with the corresponding angular compounding setting. The grating lobe reduction was calculated by $GL_{Hann} - GL_{NSI}$. As the DC offset decreased the grating lobe reduction increased for NSI. Furthermore, the grating lobes were reduced more when using incoherent summation of plane waves compared to coherent summation. The larger the number of angles used for PWI, the less the reduction in grating lobe level. These results represent a best-case scenario for high SNR. SNR in physical experiments will be much lower than achieved in the ideal simulations. To test the effects of SNR on the grating lobe levels using NSI, noise was added to the simulation data. Figure 5 shows the resulting grating lobe reduction with dc offset at 1 when noise was added to the simulated raw RF channel data and the SNR calculated from the raw unbeamformed data. The beamforming process will increase the SNR compared to unbeamformed data and, therefore, the grating lobe reduction is still higher than 0 dB in such cases. As the SNR increases the grating lobe reduction performance matches the noise free simulation results in Fig.4.

*B. Physical Experiments*

The underwater wire target images for Hann, C-NSI, IC-NSI and GCF are shown in Fig.6. GCF is implemented with fixed M equal to 2 empirically. In this figure the DC offset was set to 1.0, which produces only moderate lateral resolution improvements compared to Hann apodization. Lateral cross sections of the wire targets are shown in Fig. 7. The grating lobe reduction for C-NSI, IC-NSI and GCF with respect to Hann apodization was 20.21 dB, 23.55 dB and 11.94 dB. These results are comparable to results from the Verasonics simulation when noise was added.

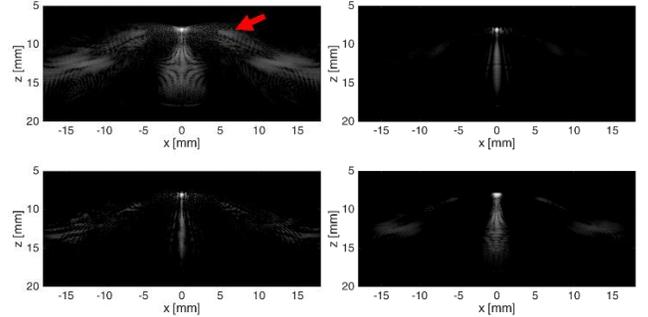

Fig.6. Underwater wire target images. Top left: Hann; Bottom left: C-NSI; Top right: IC-NSI; Bottom right: GCF. DC offset values were 1.0. Dynamic range is 60 dB for all four images.

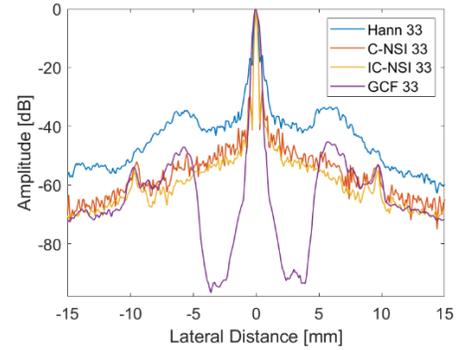

Fig. 7. Lateral profile of axially integrated power of underwater target images. DC offset values were 1.0.

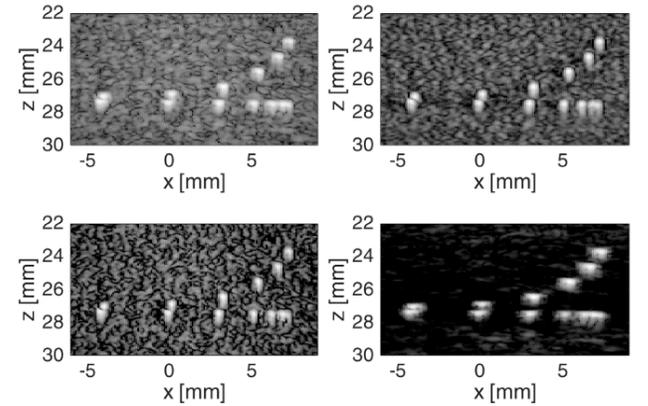

Fig.8. CIRS phantom wire target images. Top left: Hann; Bottom left: C-NSI; Top right: IC-NSI; Bottom right: GCF. DC offset values were 1.0. Dynamic range is 60 dB for all four images.

Wire targets imaged from the CIRS phantom are shown in Fig. 8. The wire targets have a speckle background, which makes it difficult to observe the grating lobe artifacts. GCF removed the speckle immediately surrounding the bright wire targets in the phantom, which results in dark spots surrounding the wire targets. This is an image artifact associated with GCF and may be undesirable for certain imaging tasks.

Contrast targets from the CIRS phantom imaged using the different techniques are shown in Fig. 9 with the NSI using a DC offset of 1.0. For this specific imaging task, GCF provided the best visual visibility with the targets and the highest contrast. The CNR values for Hann apodization, C-NSI, IC-NSI and GCF with different numbers of compounded angles measured from the marked region of CIRS phantom in Fig.9 are shown in Fig.10. Based on the metrics, IC-NSI provided better contrast than C-NSI because the speckle for C-NSI has larger variance, i.e., lower speckle SNR.

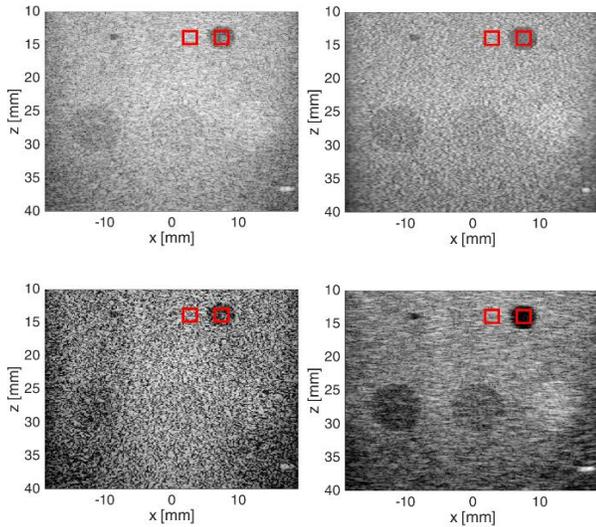

Fig.9. CIRS phantom hyperechoic and anechoic cyst images. Top left: Hann; Bottom left: C-NSI; Top right: ICNSI; Bottom right: GCF. DC offset values were 1.0. Dynamic range is 60 dB for all four images.

Tumor images from three separate rats are shown in Figs. 11, 12 and 13 for the different imaging techniques. Arrows point out the suspected grating lobe artifacts in the tumors (see tumor images created using Hann apodization). Each suspected grating lobe arose from a bright interface scatterer and were estimated to be between 36 and 41° off from the scatterer, which was the predicted location of the grating lobes based on the center frequency and pitch of the array. This supports the presence of grating lobes in the tumor images and their mitigation.

In the first rat tumor image (Fig. 11), a small bubble is present on the tumor surface laterally at about +7 mm. A grating lobe artifact (marked with an arrow) is observed from the bright bubble in the Hann apodization image inside the tumor obscuring the tissue signal underneath. The NSI images appeared to remove the grating lobe artifact completely. Other grating lobe artifacts (marked with an arrow) are present in the Hann apodization image due to the bright layer at a depth of about 14 mm near the center laterally and from a bright interface located at about 11.5 mm axially and +10 mm laterally. The grating lobe artifacts associated with these bright interfaces are removed in the NSI images. The GCF image also appears to remove some of the grating lobe artifacts but is dark and much of the tissue signal is lost degrading the image quality.

In the second rat tumor image (Fig. 12), there is a bright interface at about 7 mm axially and +7 mm laterally, which produces a grating lobe artifact on both sides of the interface in the Hann apodization image. The grating lobe artifact (marked with an arrow) is entirely removed in the NSI images. The GCF image does not entirely remove the grating lobes and they appear on both sides of the bright interface. Furthermore, the darkening of the tissue signals around bright targets is observed in the GCF images reducing their image quality.

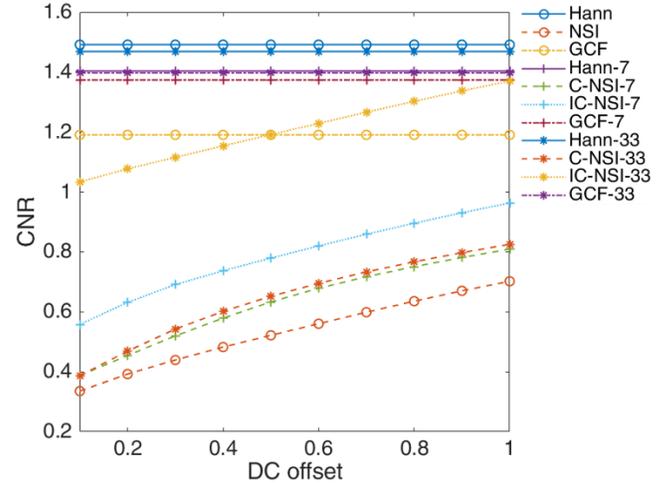

Fig.10. CNR of Hann, C-NSI, IC-NSI and GCF with different number of angle compounding from CIRS phantom scan.

The third rat tumor image (Fig. 13) shows two prevalent grating lobe artifacts from each side of the tumor generated from bright interfaces. One grating lobe artifact was generated from the interface located at 7 mm depth and -12 mm laterally. The other grating lobe artifact was generated from the interface at 11 mm depth and +7.5 mm laterally. As with the previous images, NSI appears to mitigate the grating lobe artifacts. The GCF image also eliminates the grating lobe artifact, but the image quality is degraded.

To quantify the reduction in the grating lobe levels inside the tumors, we averaged the power of the signal location at the location of the grating lobe artifacts in select regions for each tumor. The grating lobe levels were then reported in dB using the four beamforming schemes in rat tumors and are listed in Table. I. For the first tumor image, the signal was averaged for the region marked with red rectangle in the image. For the second tumor image, the signal was averaged for the region marked with red rectangle in the image. For the third tumor image, the signal was averaged for the region marked with red rectangle in the image. From Table I, C-NSI resulted in a 15-18 dB grating lobe reduction, IC-NSI resulted in 10-14 dB grating lobe reduction and GCF provided 7~8 dB grating lobe reduction compared to Hann apodization.

Table I. Integrated signal energy at the grating lobe locations for each tumor based on each beamforming approach. All values reported in dB.

|  | Hann | C-NSI | IC-NSI | GCF |
|---|---|---|---|---|
| **Tumor 1** | -39.00 | -57.71 | -53.65 | -46.05 |
| **Tumor 2** | -43.42 | -61.78 | -53.34 | -52.25 |
| **Tumor 3** | -46.37 | -61.77 | -57.68 | -53.41 |

## IV. Discussion

In this study, NSI was quantified for its ability to reduce or mitigate grating lobe artifacts for ultrasonic imaging tasks. Reducing grating lobes has increased in its importance with the advent of PWI on clinically available linear arrays that have pitches that are equal to or greater than a wavelength. Simulations and physical experiments on wire targets both indicated that the NSI approach can reduce the presence of side lobes and grating lobes. Previous studies have also revealed that NSI can reduce side lobes [4], but this current study provided evidence that NSI also reduced or eliminated grating lobes and their resulting artifacts from images.

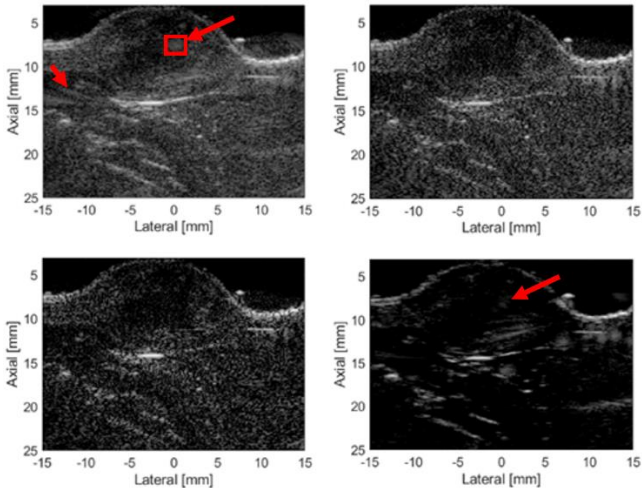

Fig.11. Rat tumor 1 images. Top left: Hann; Bottom left: C-NSI; Top right: IC-NSI; Bottom right: GCF. A DC offset of 1.0 was used for these images. Dynamic range is 60 dB for all four images.

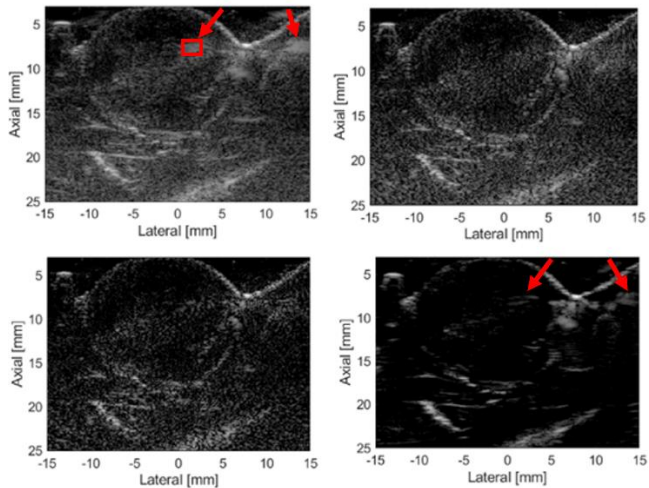

Fig.12. Rat tumor 2 images. Top left: Hann; Bottom left: C-NSI; Top right: IC-NSI; Bottom right: GCF. A DC offset of 1.0 was used for these images. Dynamic range is 60 dB for all four images.

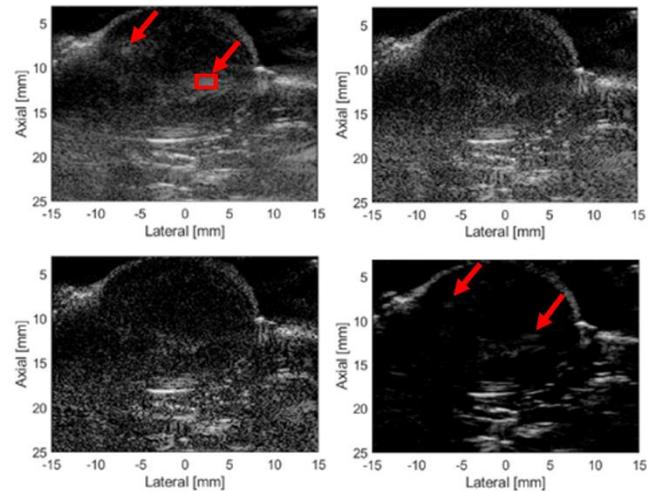

Fig.13. Rat tumor 3 images. Top left: Hann; Bottom left: C-NSI; Top right: IC-NSI; Bottom right: GCF. A DC offset of 1.0 was used for these images. Dynamic range is 60 dB for all four images.

Simulations on point scatterers indicated that NSI could reduce grating lobes by 60 to 100 dB and this reduction increased as the DC offset decreased. In NSI, the DC offset provides a tradeoff between lateral resolution and side lobe/grating lobe levels and the CNR of an image. Decreasing the DC offset results in better matching of sidelobes and grating lobes between the zero mean apodization image and the DC offset images. In addition, a lower DC offset results in a narrower NSI main beam. For an imaging task to resolve singular targets, small DC offset values are appropriate (DC ≤ 0.1). But for imaging tasks where contrast depends on low variance in the speckle within both a target and background, a low DC offset results in a high variance in the speckle leading to poor speckle SNR and CNR. Therefore, a low DC offset with NSI may not be a good option for observing contrast in ultrasound images. However, use of a larger DC offset, i.e., a DC offset of 1.0, produces speckle that is more in line with DAS and Hann apodization while still reducing sidelobes and grating lobes. Therefore, NSI imaging of tissue-mimicking phantoms and rat tumors were reported for DC offsets of 1.0 only. Images of contrast targets using NSI with small DC offsets can be found in [4].

Imaging of physical wire targets in water, i.e., a speckle-free background, revealed higher grating lobe levels using NSI than was predicted by noise-free simulations produced with the Field II. Tests of the simulations after adding noise, i.e., to more closely match the physical measurements, resulted in an increase in estimated grating lobe levels. As the SNR decreased, grating lobes levels increased because the grating lobes were not perfectly subtracted using NSI.

For the CIRS phantom images, GCF provided the best images in terms of contrast. Grating lobe artifacts were not observed in the CIRS phantom images for any of the approaches. Therefore, in a highly idealized imaging task, like imaging a test phantom, GCF produced superior results compared to NSI or Hann apodization.

However, while GCF may result in more reduction in sidelobes and grating lobes compared to NSI, there are other tradeoffs associated with GCF. For example, GCF produced shading artifacts surrounding bright scatterers in the phantom and *in vivo* images. Much of the internal structure of the rat tumors was not visible in the GCF tumor images, which created large anechoic-like regions. With additional tuning of the GCF, it is possible that the rat images could improve, but the NSI does not require tuning. The speckle of the IC-NSI in the *in vivo* rat tumor images was closest to the speckle profile of the Hann apodization images. Further, the computational load for producing GCF images is much higher compared to NSI. Therefore, these results suggest that NSI was superior at the task of imaging these rat tumors.

In comparing C-NSI and IC-NSI, two observations stand out. First, IC-NSI provided better reduction in grating lobe levels compared to C-NSI. However, these differences are hard to observe in the *in vivo* rat tumor images for C-NSI and IC-NSI. Second, higher CNR values were obtained when using IC-NSI compared to C-NSI. Coherent compounding provides better lateral resolution compared to incoherent compounding, however, incoherent compounding has been used for decades to smooth speckle and increase CNR [45]. This reduction in CNR is related to the decreased speckle SNR produced by the nonlinear processing used in NSI. The speckle SNR was calculated for a homogeneous region in the CIRS phantom for the Hann, C-NSI and IC-NSI. The speckle SNR for the Hann images, C-NSI and IC-NSI were 1.69, 0.97 and 1.42, respectively. This indicates that C-NSI was far from producing fully developed speckle (value of 1.91). With the large reduction in grating lobe levels afforded by NSI using either C-NSI or IC-NSI, the superior CNR provided by IC-NSI suggests IC-NSI is the better choice for general ultrasonic imaging tasks compared to C-NSI when using PWI with multiple transmission angles. However, there are more difficult imaging situations, such as multipath, reverberation and large sound speed inhomogeneities in real-world imaging tasks where the performance of NSI should be evaluated.

## V. Conclusion

In this work, the NSI beamforming technique was evaluated for reducing the grating lobes associated with PWI using clinically available clinical linear arrays. The presence of grating lobes produces increased clutter in ultrasonic images, which in turn reduces image quality. Furthermore, clutter may obscure important information in images. Two NSI approaches, C-NSI and IC-NSI, were evaluated for specific imaging tasks. For general imaging *in vivo*, the results suggest that IC-NSI (DC offset of 1.0) provided the best performance in terms of spatial resolution, contrast and clutter reduction from grating lobes.

## Acknowledgment

The authors wish to acknowledge the assistance of Kendall Junger, Divya Nambiar and Erin Kinaci with cell culture.